%% file: DecisionReferrals_main.tex
\documentclass[letterpaper, 10pt, journal]{IEEEtran} 
\usepackage[utf8,utf8x]{inputenc}
\usepackage{amsmath,amscd,amssymb,amsgen,amsfonts,amsbsy}
\usepackage{mathrsfs}
\usepackage{textcomp}
\usepackage{float}
\usepackage{latexsym}
\usepackage{mathtools}
\usepackage{breqn}
\usepackage{tikz}
\usetikzlibrary{automata,arrows,positioning,calc}
\usepackage{authblk}
\usepackage{url}
\usepackage{cite}
\usepackage{algorithmic}
\usepackage[ruled,lined]{algorithm2e}
\SetCommentSty{itshape}  
\usepackage{graphicx}
\usepackage{subfigure}
\usepackage{caption}
\usepackage{subcaption}
\usepackage{enumitem}

\usepackage{pbalance}

\newcommand{\tup}[1]{\mathit{#1}}
\newcommand{\mcl}[1]{\mathcal{#1}}

\newcommand{\comments}[1]{}

\newcommand{\Hyp}{\mathcal{H}}

\newcommand{\Qfun}{\mathcal{Q}}

\newtheorem{theorem}{Theorem}

\newtheorem{assumption}{Assumption}
\newtheorem{problem}{Problem}

\newtheorem{example}{Example}

\newcommand{\autom}{a}  
\newcommand{\human}{h}

\title{Task load dependent decision referrals for joint binary classification in human-automation teams}

\author{Kesav Kaza, Jerome Le Ny and Aditya Mahajan%
\thanks{This work was supported by the Canadian Department of National Defence under the Innovation for Defence Excellence and Security (IDEaS) Micro-Net program grant CFPMN2037, and by FRQ-NT grant 343486.}
\thanks{A preliminary version of this paper was presented at the IEEE Conference on Decision and Control, December 2021. The current version presents an experimental study with 
human subjects for model parameter estimation and statistical validation of the proposed referral 
schemes in addition to updated analysis, simulation experiments and literature survey.}
\thanks{Part of this research involved human subjects. The experimental procedures were reviewed and approved 
by Polytechnique Montreal's research ethics committee under project CER-2122-43-D.}
\thanks{Kesav Kaza was with Polytechnique Montreal, Canada at the time of this work.
Jerome Le Ny is with the Department of Electrical Engineering, Polytechnique Montreal, Canada. Aditya Mahajan is with the Department of Electrical and Computer Engineering, McGill University, Canada. 
All authors were affiliated with GERAD, Montreal, Canada at the time of this work.
{\tt\small kkaza@uottawa.ca,\tt\small jerome.le-ny@polymtl.ca},
{\tt\small aditya.mahajan@mcgill.ca}} }

\begin{document}

\maketitle
\begin{abstract}
We consider the problem of optimal decision referrals in human-automation teams performing binary classification tasks. The automation, which includes a pre-trained classifier, observes data for a batch of independent tasks, analyzes them, and may refer a subset of tasks to a human operator for fresh and final analysis. Our key modeling assumption is that human performance degrades with task load. We model the problem of choosing which tasks to refer as a stochastic optimization problem and show that, for a given task load, it is optimal to myopically refer tasks that yield the largest reduction in expected cost, conditional on the observed data. 
This provides a ranking scheme and a policy to determine the optimal set of tasks for referral.
We evaluate this policy
against a baseline through an experimental study with 
human participants. Using a radar screen simulator, participants made binary target classification decisions under time constraint. They were guided by a decision rule provided to them, but were still prone to errors under time pressure. An initial experiment estimated human performance model parameters, while a second experiment compared two referral policies. Results show statistically significant gains for the proposed optimal referral policy 
over a blind policy that determines referrals using the automation and human-performance models 
but not based on the observed data.
\end{abstract}

\input{texFiles/introduction}

\input{texFiles/systemModel}

\input{texFiles/results}

\input{texFiles/simulations}

\input{texFiles/experiments_human}

\section{Conclusion}\label{section: conclusion}

A decision referral problem is formulated for a human-automation team jointly 
performing binary classification tasks with decision costs. 
The automation decides which tasks should be referred to the human for final 
classification decisions, after performing a first analysis of the data. 
An algorithm is presented to find the optimal referral decisions. 
The proposed 
method only requires the true and false positive rates of the 
human operator as a function of task load, 
rather than a model of the human decision making process. 
Simulations and experiments were conducted to compare the optimal referral policy (OA),
a constant task load version of OA (SA), and a blind allocation scheme (BA) that does not analyze the tasks but computes an optimal task load 
using the human's and the automation's performance models. 
Results show that OA performs significantly better than BA,
and that SA may 
represent a good alternative to OA in applications where frequent changes in human task load are undesirable.
Future 
work could extend this study to
consider 
sequential decision making applications and the impact of
other human factors such as fatigue and trust in automation.


\bibliographystyle{IEEEtran}
\bibliography{IEEEabrv,HRteams}

\end{document}

%% file: texFiles/introduction.tex
\section{Introduction}

There is a growing interest in developing
collaborative systems where automated agents and human operators work together on tasks such as monitoring an environment 
or an industrial control system \cite{Kumar:SMC2020:HRCsurvey},
manipulating objects \cite{Music:ARC17:controlsharing}, identifying dynamic threats or searching for objects 
in military and public safety applications \cite{Wang:TCST16:search}. 
Furthermore, to support the design of automated decision support systems (DSS), 
considerable research has been done on modeling 
and assessing 
the performance of human-AI systems \cite{Lai:2021:towards-human-ai, Reverberi:Nature2022:human-ai-medical, deArteaga:2020:case-for-HILP, Amazu:2024:experimentData,Grover:2020:RADAR-DSS}.
Ensuring 
effective collaboration between an automated agent and a human operator 
requires careful design of task allocation strategies that account for the decision-making hierarchy. 
In this paper, 
we develop and evaluate such a strategy 
for binary classification tasks performed by a human operator with the help of a DSS. 

There is extensive literature on distributed and collaborative decision-making, including distributed hypothesis testing and filtering \cite{Tsitsiklis:SSP93:detection,Tartakovsky:book14:sequential,Hyun:CDC-ECC11:nestedclassification}.
Moreover,
a growing body of work on 
human-in-the-loop machine learning and reinforcement learning explores
human feedback for minimizing errors, improving generalization 
and test-time adaptation of the learned models~\cite{Mosqueira:2023:HITL-ML, Retzlaff:2024:HITL-RL, Peng:2023:HITL-TTA}.
%
%
However, this literature does not consider the impact of various human factors 
such as cognitive workload, fatigue, trust or belief in the automation's capability,
which have significant influence
on the joint performance of a human-machine system \cite{Chen:HRI18:planning,Dubois:SMC20:trust,Gao:TSMC06:DFT,Wickens:book15:humanperf}.



In human-automation teams, tasks may be complex and demand different kinds of mental and physical resources such as attention,  memory, motor functions, etc.~\cite{Wickens:book15:humanperf}. Cognitive workload (CW) is an important factor that captures these 
features and
much effort has been devoted to its quantification and measurement~\cite{Wickens:book15:humanperf,Sheridan:MentalWorkload79:Definitions,Tulga:TSMC80:workload, Hart:HumanFactorsErgonorics06:NasaTLX}. 
%
%
However, for simple classification tasks where the human operators use 
pre-determined decision rules for classification, it may be reasonably assumed that CW 
has a monotonic relationship with task load, defined simply as the number of tasks 
assigned to an operator in a fixed time period. Hence, task load can be used as a handle on CW~\cite{Tulga:TSMC80:workload, Cummings:IEEE14:manMachine,Cummings:Handbook09:collaborative}. 

Some literature on task load dependent task allocation to humans 
assumes that tasks are buffered in a queue and served sequentially by the human who has a utilization dependent service rate \cite{Jog:arxiv2021:udc}. Two classes of problems in this context relate to task release policies stabilizing the queues \cite{Savla:TAC10:queue,Srivastava:ACC12:taskreleasequeue,Lin:arxiv2021:stabilizing}, and optimal time or attention allocation policies for human operators \cite{Srivastava:AMC11:queues}.
However, in these papers the nature of the tasks is abstract, and the solutions proposed do not typically apply to collaborative decision-making problems. 

The literature on task allocation for collaborative decision-making may broadly be categorized based on the behavior of the human operator as `interventionist' or `non-interventionist'. 
The operator might decide to intervene during the AI's task as in \cite{Chen:THRI20:trustPOMDP}, or does not intervene 
but may 
review or redo some tasks completed by the automation as in \cite{Li:TASE24:TrustAware, Dubois:SMC20:trust}. 
Task allocation methods considering human factors can also be characterized as `blind' or `informed' depending if they find an optimal task load distribution for the team a priori 
or based on the uncertainty involved in the individual tasks that might require human review. 

In this paper, we propose
a hierarchical decision framework for human-automation teams working on classification tasks.  The  
decision architecture allows joint optimization of the team performance by focusing the limited cognitive resources 
of the human operator on the tasks that 
the automation determines to be the most beneficial to refer, based on its observed data.
A collaborative decision-making architecture close to our work is considered in \cite{Hyun:TC15:nestedclassification,Chitalia:ACC14:revisit}, with a nested hierarchical decision structure 
where a human operator either chooses to or is asked to review a decision.
However, these studies focus on generic tasks in a steady-state environment, where the workload depends on the size of the ``confusion region'' and can be interpreted as the probability of referring a task to the human. Furthermore, the analysis 
assumes specific probabilistic models (e.g., Gaussian 
or binary observations) and a specific task referral heuristic. 

The main contributions of our paper are as follows:
\begin{enumerate}[leftmargin=*]
    \item We show that the optimal task referral policy for our framework has a structure that makes it easy to implement. In particular, we define a task load dependent referral index for each task observed by the automation and prove that it is optimal to allocate tasks with the $w$-highest referral indices to the human operator, where $w$ is the task load. The overall optimal referral policy can then be determined by searching over all feasible values of $w$.

    \item A salient feature of the referral indices is that they can be computed based on the human operator's true positive and false positive rates as a function of task load of the human operator. These functions can be empirically estimated through experiments with human subjects, making the proposed policy practical for real-world implementation.
    
    \item Another salient feature of the optimal policy is that the task load depends on the batch of tasks observed by the automation. We present two variants of the optimal policy, which keep the task load constant. Such variants may be used when varying the human's task load  is undesirable. We present a detailed simulation study that compares the performance of the optimal allocation policy with constant task load policies.

    \item  We validate the performance of the proposed optimal referral policy against a baseline in a semi-realistic setup  through an experiment with human subjects. This setup simulates the classification of mobile  objects (labeled as ``Hostile'' or ``Non-Hostile'') tracked on a radar screen.
\end{enumerate}

The rest of the paper is organized as follows. In Section~\ref{section: model}, we present
the system model and formulate the problem of optimal decision referrals. 
In Section~\ref{section: results}, we present an algorithm to obtain the optimal decision referral policy. 
In Section~\ref{section: simulations}, we present a simulation study comparing the performance 
of the proposed algorithm with baseline allocation schemes and also study the impact of errors 
in the model parameters. In Section~\ref{sec:humans}, we describe the experimental study 
with human participants. Finally, we conclude in Section~\ref{section: conclusion}.

\paragraph*{Notation} The notation $\mathbb{P}(\cdot)$ denotes the probability of an event, $X \sim 
\mathcal{N}(\mu, \sigma^2)$ means that $X$ is a random variable distributed according to the Gaussian distribution with mean $\mu$ and variance $\sigma^2$, and $X \sim U(a,b)$ means that $X$ is a random variable distributed according to the uniform distribution over the interval $[a,b]$. We also use $X_{1:K}$ as a shorthand notation for the sequence $\{X_1,\ldots,X_K\}$.

%% file: texFiles/systemModel.tex
\section{System Model and Problem Formulation}
\label{section: model}

Consider an automation and a human operator jointly classifying a batch of $K$ independent binary classification tasks, {indexed by} $\mathcal{K}\coloneqq\{1,\ldots,K\}$.
Each task $k\in\mathcal{K}$ has a random binary state $H_k\in\{\Hyp_0,\Hyp_1\}$.
The states $H_{1:K}$ are independent and identically distributed, where $\pi_{0}=\mathbb{P}(H_k=\Hyp_0)$ and $\pi_1 = \mathbb{P}(H_k = \Hyp_1)$ are known. 

For every task $k\in\mathcal{K}$, the automation and the human operator respectively receive observations $Y^{\autom}_{k}\in \mathcal{Y}^\autom$ and $Y^{\human}_k \in \mathcal{Y}^\human$ 
that are conditionally independent given $H_k$. 
Moreover, for $m \in \{\autom, \human\}$, the conditional distribution of $Y^m_k$ given $H_k$ is identical for all tasks $k \in \mathcal{K}$.
The specific details of the observation model are provided in the next section.

The system operates as follows. First, the automation sees the observations $Y^{\autom}_{1:K}$ of the entire batch and, for each task $k \in \mathcal{K},$ decides to either classify the task as $\Hyp_0$ or $\Hyp_1$ or to refer (i.e., transfer) the task to the human operator. Let $\mathcal{N}\subseteq \mathcal{K}$ denote the indices of the tasks referred to the human opeator. 
The human operator then sees the observations $\{Y^{\human}_n\}_{n\in\mathcal{N}}$ of the tasks referred to her and classifies each of the referred task~$n \in \mathcal{N}$ as $\Hyp_0$ or~$\Hyp_1$.

\subsection{Observation Models}

We assume that the automation has a known fixed observation model {$P^\autom(\cdot | \mathcal H_i)$, {which for each} $i \in \{0,1\}$ defines a probability distribution on $\mathcal Y^\autom$}. {Consequently, the joint probability of the automation's observations is given by}
\[
{\mathbb{P}(Y^{\autom}_{1:K})
= \prod\limits_{k\in\mathcal{K}}\sum\limits_{i\in\{0,1\}}\pi_{i}P^{\autom}(Y^{\autom}_k|\Hyp_i).}
\] 
In contrast, the human operator's observation model depends on the task load $w \coloneqq |\mathcal{N}|$, 
which is defined as the number  of tasks referred to her by the automation. {Let $\mathcal{W} \subseteq \{0, 1, \dots, K\}$ denote the set of possible task loads.\footnote{In principle it is possible to always take $\mathcal{W} = \{0, 1, \dots, K\}$. However, in practice, we may want to avoid low task loads and restrict $\mathcal{W}$ to $\{K_{\min}, \dots, K\}$.}}
Then, for each $w \in \mathcal{W}$, the observation model of the human operator is given by $P^{\human}( \cdot |\Hyp_i,w)$, which for each $i \in \{0,1\}$ defines a probability distribution on $\mathcal Y^\human$. Thus, the joint probability of the human operator's observations given a task load $w$ is
\[
\mathbb{P}(\{Y^{\human}_n\}_{n\in\mathcal{N}} | w) 
= \prod\limits_{n\in\mathcal{N}}\sum_{i\in\{0,1\}}\pi_{i} P^{\human}(Y^{\human}_n|\Hyp_i,w).
\] 
This task load dependent observation model captures a degradation of the human operator's observations with an increase in task load, since the operator must dedicate less time or 
cognitive resources to each individual task.

In general, the human's observation would become noisier with increasing task load. Furthermore, 
for the decision referral problem to be non-trivial, it must  not be universally better to allocate all the tasks to the human operator or to the automation. Thus, in general we would expect that the observations of the automation are noisier than the human at low task load whereas the opposite is true at high task load.

We now present an example of an analytical Gaussian observation model for the automation and the human operator that captures these trade-offs. Gaussian observation models are used in the literature to represent decision making  under uncertainty~\cite{Wickens:book02:SDT} and are also relevant in applications involving high dimensional  inputs (such as images) classified using neural networks, where the weights and  input to  the output layer may be assumed to be Gaussian by applying a mean field 
approximation \cite{Schoenholz:2016:DeepInfoProp}.

\begin{example}[Gaussian Observation Models]
\label{ex: Gaussian models general}
In this model the observations of the automation are given as follows. 
For any $k \in \mathcal{K}$,
\begin{align*}
\text{under $\Hyp_0$}, \quad  Y^{\autom}_k &\sim \mathcal{N}(0,\sigma^2_{\autom}), \\
\text{under $\Hyp_1$}, \quad  Y^{\autom}_k &\sim \mathcal{N}(d_\circ,\sigma^2_{\autom}),
\end{align*}
where $d_\circ$ and $\sigma_0$ are constants. 
In contrast, the observations of the human operator are given as follows. 
For any $n \in \mathcal{N}$, 
\begin{align*}
\text{under $\Hyp_0$}, \quad Y^{\human}_n &\sim \mathcal{N}(0,\sigma^2_{\human}(w)), \\
\text{under $\Hyp_1$}, \quad Y^{\human}_n &\sim \mathcal{N}(\mu_{\human}(w)),\sigma^2_{\human}(w)),
\end{align*}
where the mean $\mu_{\human}(w)$ and the variance $\sigma^2_{\human}(w)$ depend on the task load $w \in \mathcal{W}$. Two examples of such dependence are:
\begin{enumerate}[leftmargin=*]
    \item $\mu_{\human}(w) = (1 - w/K) \mu_\circ$ and $\sigma^2_{\human}(w)= \sigma^2_{\circ}$, where $\mu_{\circ}$ and $\sigma^2_{\circ}$ are predefined constants such that 
    $0 = \mu_{\human}(K)< d_\circ < \mu_{\human}(1) = (1-1/K)\mu_{\circ}$ 
    and  $\sigma^2_{\circ} \leq \sigma^2_a$.
    Here, as the task load increases the conditional densities under both hypotheses 
    move closer, which degrades binary classification performance. 
    \item $\mu_{\human}(w) = d_\circ$,  
    $\sigma^2_{\human}(w) = (1+w/K)\sigma^2_{\circ}$, where $\sigma^2_{\circ}$ is a predefined constant such that $(1+1/K)\sigma_{\circ}^2 < \sigma_\autom^2 < {2} \sigma_{\circ}^2$. Here, as the task load increases, the observation variance under both hypothesis increases and eventually becomes larger than the observation variance for the automation.
\end{enumerate}
\end{example}

\subsection{Human Decision Model}
\label{section: human decision-making}

We make the following assumption on the decision making process of the human. 
\begin{assumption}  \label{assumption: non-interacting decisions}
For each task $n \in \mathcal{N}$, the human operator decides between
$\Hyp_0$ and $\Hyp_1$ based only on her observation $Y^{\human}_n$. 
In particular, the human operator does not have access to the automation's observation $Y^{\autom}_n$, 
and in her decision does not account for the fact that the automation referred the task to her 
after observing 
$Y^{\autom}_{1:K}.$
\end{assumption}

Assumption \ref{assumption: non-interacting decisions} can be justified based
on the limited time that the operator has to make a decision.
Note that $Y^{\autom}_{n}$ and $Y^{\human}_n$ represent high-level features 
abstracted respectively by the automation's signal processing pipeline and 
the human's cognitive process, on which the final decisions {are} based. {Thus, $Y^{\autom}_n$ and $Y^{\human}_n$ would differ} 
even in situations where the human operator and the automation observe the same raw data
(for example, a picture or a text message). {This is especially true in scenarios where the automation uses neural network models}, which are often opaque and not easily interpretable by humans.

Let $D_k$ denote the final classification decision on task $k \in \mathcal{K}$,  made either by the automation or the human operator. Let $P^{\human}_{\tup{tp}}$, $P^{\human}_{\tup{fp}}$: $\mathcal{W}\rightarrow [0,1]$ be  the true positive rate (TPR) and false positive rate (FPR) of the human, depending on task load, i.e., for all $n \in \mathcal{N}$
\begin{subequations}    \label{eq: S2 decision probabilities}
\begin{align}
\mathbb{P}(D^{\human}_n=\Hyp_1|{H}_n=\Hyp_1,w)
&= P^{\human}_{\tup{tp}}(w),
\\
\mathbb{P}(D^{\human}_n=\Hyp_1|{H}_n=\Hyp_0,w) &= P^{\human}_{\tup{fp}}(w). 
\end{align}
\end{subequations}
In practice, the TPR and FPR functions in \eqref{eq: S2 decision probabilities} can be obtained 
through calibration experiments with the human operator \cite{Dubois:SMC20:trust,Wickens:book02:SDT}, 
as described in Section \ref{section: model fitting}.
As we show later, these functions capture all the information about the human operator sufficient for the automation to compute its optimal referral policy, and in particular it does not need to model the decision-making process of the human.
Nevertheless, we will also discuss examples where these functions may be obtained from first-principle reasoning for our simulations in Section~\ref{section: simulations}.

\subsection{Decision Costs and Optimization Problem}

We assume that for any task $k \in \mathcal{K}$, the cost of making a terminal decision $D_k$ when the true hypothesis is $H_k$ is captured by a cost function $C(D_k, H_k)$. 
For the ease of notation, we define
\begin{subequations}
    \label{eq: decision_costs}
    \begin{align}
    c_{\tup{tp}} &\coloneqq C(\Hyp_1, \Hyp_1), &
    c_{\tup{fp}} &\coloneqq C(\Hyp_1, \Hyp_0), \\
    c_{\tup{tn}} &\coloneqq C(\Hyp_0, \Hyp_0), &
    c_{\tup{fn}} &\coloneqq C(\Hyp_0, \Hyp_1).
    \end{align}
\end{subequations}
In addition, we assume that the system incurs a cost $c_r$ for each task referred to the human.

{After} the automation has observed a batch $Y^a_{1:K}$, the total expected cost incurred by the system from the point of view of the automation depends on the automation's posterior belief $p^a_{1:K}$, which is defined as
\begin{equation}    \label{eq: automation posterior}
p^{\autom}_{k}\coloneqq \mathbb{P}(H_k=\Hyp_1|Y^{\autom}_k), \quad k\in\mathcal{K}.
\end{equation}
The posterior beliefs can be computed using Bayes rules as follows:
\[
  p^{\autom}_k = \frac{ \pi_1 P^{\autom}(Y^{\autom}_k | \Hyp_1) }
  {\pi_0 P^{\autom}(Y^{\autom}_k | \Hyp_0) + \pi_1 P^{\autom}(Y^{\autom}_k | \Hyp_1) }.
\]

From the point of view of the automation, for tasks $k \in \mathcal{K} \setminus \mathcal{N}$ where the automation makes a terminal decision,  the system incurs an expected cost
\begin{align}   \label{eq: automation decision expected cost}
{\Gamma^\autom}({p^{\autom}_{k},D_k}) = (1-p^{\autom}_{k}) C(D_k,{\mcl H_0}) 
+ p^{\autom}_{k} C(D_k,{\mcl H_1}).
\end{align}
For tasks {$n \in \mathcal{N}$ that the automation} refers to the human operator, the automation does not know the  final decisions $\{D_n\}_{n \in \mathcal{N}}$ and needs to form a posterior belief on them. Thus, from the point of view of the automation, the expected cost of referring a task {$n \in \mathcal N$} to the human operator is
\begin{align}   \label{eq: human classification cost per task}
\Gamma^\human(p^{\autom}_{n},w) &= c_r + (1-p^{\autom}_{n}) \left[P^{\human}_{\tup{fp}}(w)c_{\tup{fp}} + (1-P^{\human}_{\tup{fp}}(w))c_{\tup{tn}} \right] \nonumber\\
&\quad + p^{\autom}_{n}\left[P^{\human}_{\tup{tp}}(w) c_{\tup{tp}} + (1-P^{\human}_{\tup{tp}}(w))c_{\tup{fn}} \right].
\end{align}
Combining both the above expected costs, we get that the expected total system cost from the point of view of the automation is
\begin{align}  
\hskip 1em & \hskip -1em 
J(\mcl{N},\{D_k\}_{k\in\mcl{K}\setminus \mcl{N}}, p^{\autom}_{1:K}) \notag \\
&= \sum_{k\in\mcl{K}\setminus \mcl{N}} \Gamma^{\autom}(p^{\autom}_{k},D_k) + \sum_{n \in \mcl N} \Gamma^\human(p^{\autom}_{n},|\mcl{N}|). 
\label{eq:team_cost}
\end{align}
The observations $Y^{\autom}_{1:K}$ are used to form  the posteriors $p^{\autom}_{1:K}$ given by \eqref{eq: automation posterior} and can then be discarded,  which explains the choice of variables in $J$.
We are then interested in the following problem.
\begin{problem}
\label{problem_def}
Given the posterior beliefs $p^{\autom}_{1:K}$ of the automation and the  human operator's TPR and FPR functions $P^{\human}_{\tup{tp}}$, $P^{\human}_{\tup{fp}}$ {defined in \eqref{eq: S2 decision probabilities}},  determine $\mcl{N}$ and $\{D_k\}_{\mcl{K}\setminus\mcl{N}}$ so as to maximize  $J(\mcl{N},\{D_k\}_{k\in\mcl{K}\setminus \mcl{N}}, {p^{\autom}_{1:K}})$ given by~\eqref{eq:team_cost}.
\end{problem}

%% file: texFiles/results.tex
\section{Optimal Referral Policy}
\label{section: results}

In this section, we present a solution framework to efficiently solve Problem~\ref{problem_def}. Our solution proceeds in three steps. In step one, we consider the subproblem where the set $\mathcal{N}$ of tasks referred to the human operator are fixed and identify the optimal decision policy of the automation. In step two, we consider the subproblem where the number of tasks $|\mathcal{N}|$ to be referred to the human operator is fixed and identify a qualitative structure of the optimal referral policy (Theorem~\ref{thm:fixed_w}). In step three, we find the overall optimal referral policy by searching over all values of task load $|\mathcal{N}|$ and using the solution of step one to find the best solution in each case (Theorem~\ref{thm:main}). 


\begin{figure}[!t]
\centering
\includegraphics[width=0.85\columnwidth]{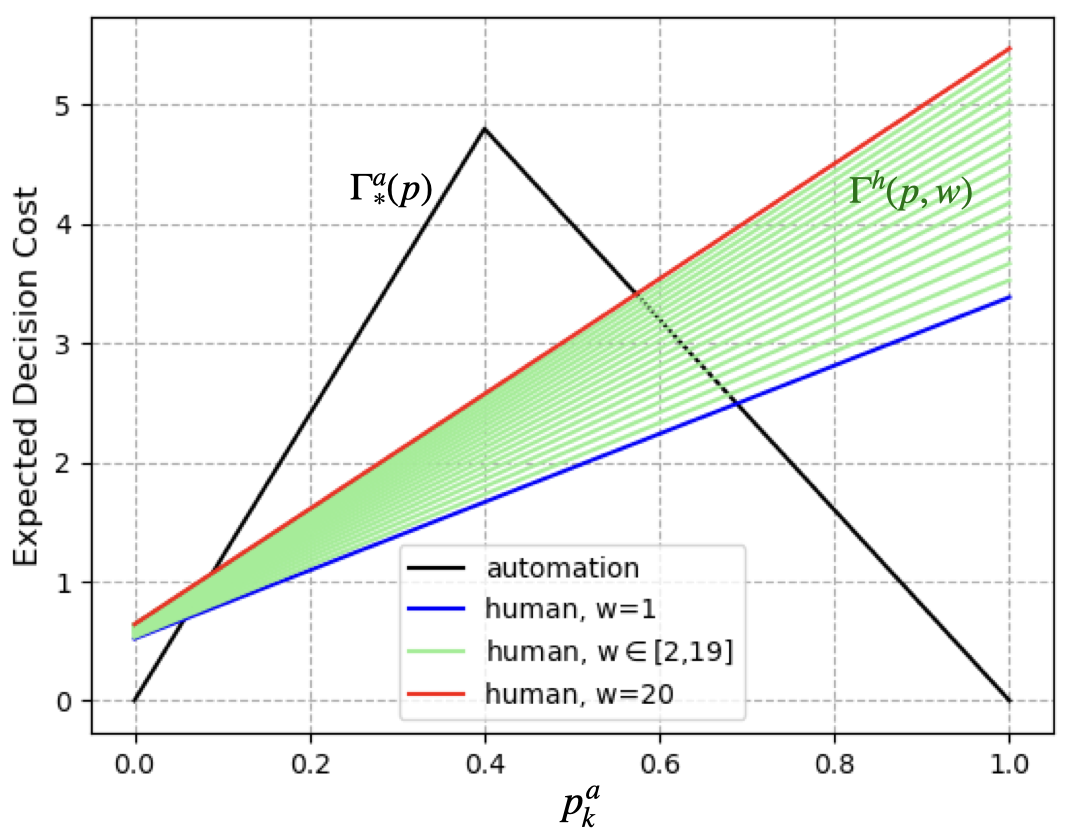}
\caption{Illustration of the classification costs of the automation, $\Gamma^{\autom}_*(p)$, and the human, $\Gamma^{\human}(p,w)$, for different task loads as a function of the posterior belief about a task. The referral index is the difference between these two curves.}
\label{fig:Ex1_C1_EC2_w}
\end{figure}

\subsection{Step one: automation's optimal policy for fixed $\mathcal{N}$}

Suppose the set $\mathcal{N}$ is fixed. For each task $k \in \mathcal{K}\setminus\mathcal{N}$, the automation has to make a terminal decision $D_k \in \{\Hyp_0, \Hyp_1\}$ to minimize $\sum_{k \in \mathcal{K}\setminus\mathcal{N}} \Gamma^{\autom}(p^\autom_k, D_k)$. Each of these decisions are decoupled. Therefore, for each $k \in \mathcal{K}\setminus\mathcal{N}$ the automation should choose
\begin{align}
   D_k &= \arg\min_{D_k \in \{\Hyp_0, \Hyp_1\}} \Gamma^{\autom}(p^{\autom}_k, D_k)
   \notag \\
   &= \begin{cases}
        \Hyp_0, & \text{if  } \Gamma^{\autom}(p^{\autom}_{k},\Hyp_0)\leq \Gamma^{\autom}(p^{\autom}_{k},\Hyp_1),\\
        \Hyp_1, & \text{ otherwise}.
    \end{cases}
   \label{eqn:auto_Bayes_decision}
\end{align}
Furthermore, the corresponding optimal cost is given by 
\begin{equation}   \label{eq: optimal automation cost per task}
\Gamma^\autom_{*}(p^{\autom}_{k}) \coloneqq  \min\lbrace \Gamma^{\autom}(p^{\autom}_{k},\Hyp_0),\Gamma^{\autom}(p^{\autom}_{k},\Hyp_1) \rbrace.
\end{equation}

\subsection{Step two: automation's optimal policy for fixed $|\mathcal{N}|$}

Suppose now $w = |\mathcal{N}|$ is fixed and the automation has to decide which $w$ tasks to refer to the human operator. For the tasks not referred to the human operator, the terminal decisions are made as per~\eqref{eqn:auto_Bayes_decision}, incurring a cost given by~\eqref{eq: optimal automation cost per task}. The key idea of our solution is to define what we call a \emph{referral index} $R \colon [0, 1] \times \mathcal{W} \to \mathbb{R}$, as follows:
\begin{equation} \label{eqn:G_index}
    R(p,w) \coloneqq \Gamma^{\autom}_*(p) - \Gamma^{\human}(p, w), \quad
    p \in [0,1], w \in \mathcal{W}.
\end{equation}
Note that $\Gamma^{\autom}_*(p)$ defined in~\eqref{eq: optimal automation cost per task} is piecewise linear in $p$ and, for a fixed $w$, $\Gamma^{\human}(p,w)$ defined in~\eqref{eq: human classification cost per task} is linear in $p$. Thus, for a fixed $w$, $R(p,w)$ is piecewise linear in $p$. Furthermore, for a fixed $p$, we expect $\Gamma^{\human}(p,w)$ to be weakly increasing in $w$; so $R(p,w)$ would be weakly decreasing in $w$.

As an illustration of the computation of the referral index, consider an example with batch size $K = 20$, prior probability $\pi_0 = 0.8$ where the observation model is as given in Example~\ref{ex: Gaussian models general} - case 1,
with $d_\circ = 3$, $\sigma_\circ = 1.8$, $\sigma_a = 1.6$, decision costs are $c_{\tup{tp}}=c_{\tup{tn}}=0$, $c_{\tup{fp}}=8$, $c_{\tup{fn}}=12$, and referral cost is {$c_r = 0.5$.} For this model, $\Gamma^{\autom}_*(p)$ is given by the black curve in Fig.~\ref{fig:Ex1_C1_EC2_w} while $\Gamma^{\human}(p,w)$ is shown by the colored lines for different values of $w$. The referral index is the difference between these two curves.

Our main result is to show that the optimal choice of which $w$ tasks to refer to the human can be determined from the referral indices. 

\begin{theorem} \label{thm:fixed_w}
To minimize  $J(\mcl{N},\{D_k\}_{k\in\mcl{K}\setminus \mcl{N}}, p^{\autom}_{1:K})$ in \eqref{eq:team_cost}
for a given task load $w = \vert\mcl{N}\vert$ and posteriors $p^a_{1:K}$, it is optimal for the automation to refer 
the $w$ tasks with largest referral indices among $\{R(p^a_{k},w)\}_{k=1}^K$ and to take decisions  for the other tasks according to \eqref{eqn:auto_Bayes_decision}.
\end{theorem}
\begin{IEEEproof}
By definition of $\Gamma^{\autom}_*(p)$ and the referral indices, we have that
{\small{
\begin{align}
J(\mcl{N},\{D_k\}_{k\in\mcl{K}\setminus \mcl{N}},p^{\autom}_{1:K}) &\stackrel{(a)}{\ge} \sum\limits_{k\in\mcl{K}\setminus\mcl{N}} \Gamma^{\autom}_*(p^{\autom}_{k}) + \sum\limits_{n\in\mcl{N}} \Gamma_\human(p^{\autom}_{n},\vert\mcl{N}\vert) \notag \\
&= \sum\limits_{k\in\mcl{K}} \Gamma^{\autom}_*(p^{\autom}_{k}) -  
\sum\limits_{n\in\mcl{N}}R(p^{\autom}_{n},\vert\mcl{N}\vert),
\label{eq:cost_func_team}
\end{align} }}
with equality achieved in $(a)$ when the decisions 
$\{D_k\}_{k\in\mcl{K}\setminus\mcl{N}}$ are chosen according to \eqref{eqn:auto_Bayes_decision}. 
Now note that the first term in \eqref{eq:cost_func_team} is constant.
So for $\vert\mcl{N}\vert$ fixed, \eqref{eq:cost_func_team}
is minimized by choosing $\mcl{N}$ to maximize 
 $\sum_{n\in\mcl{N}} R(p^{\autom}_{n},\vert\mcl{N}\vert)$,
which is indeed achieved by selecting the tasks with highest referral indices.
\end{IEEEproof}

\subsection{Step three: automation's overall optimal policy}

When the task load $w = |\mathcal{N}|$ is not fixed, we can consider all possible values of $w \in \mathcal{W}$ to find the $w$ with the minimum overall cost. An algorithm describing the complete procedure is shown in Algorithm~\ref{algo:optworkload}. The input to the algorithm is the vector of posterior beliefs $p^\autom_{1:K}$ and the output is a set $\mathcal N$ of tasks to be referred to the human.

\begin{algorithm}[htbp]
\KwIn{Posterior beliefs $p^{\autom}_{1:K}$}
\KwOut{Set $\mathcal{N}$ of tasks reffered to the human}
Initialize $\mathcal N = \emptyset$, $\delta^* = -\infty$ \\
\For{$w\in\mathcal{W}$}
{
Compute referral indices $\{R(p^{\autom}_k, w)\}_{k \in \mathcal{K}}$ using~\eqref{eqn:G_index}. \\
$\mathcal{N}_w = \text{$w$ tasks with top indices in } \{R(p^{\autom}_k, w)\}_{k \in \mathcal{K}}$. \\
Compute cost reduction $\Delta(w) = \sum_{n \in \mathcal{N}_w} R(p^{\autom}_n, w)$. \\
\If{$\Delta(w) > \delta^*$} {
    $\delta^* \gets \Delta(w)$ \\
    $\mathcal N \gets \mathcal N_w$ \\
   }
}
\Return{$\mathcal N$}
\caption{Optimal task referral}
\label{algo:optworkload}
\end{algorithm}

\begin{theorem}\label{thm:main}
   The optimal solution of Problem~\ref{problem_def} is to compute $\mathcal{N}$ via Algorithm~\ref{algo:optworkload}, refer these tasks to the human, and apply terminal decisions $D_k$ as per~\eqref{eqn:auto_Bayes_decision} for the remaining tasks $k \in \mathcal{K}\setminus\mathcal{N}$.
\end{theorem}
\begin{IEEEproof}
Recall that it was shown in the proof of Theorem~\ref{thm:fixed_w} that for a fixed task load $w$, 
\[
    J(\mathcal{N}, \{D_k\}_{k \in \mathcal{K}}, p^{\autom}_{1:K}) \ge 
    \sum_{k \in \mathcal{K}} \Gamma^{\autom}_*(p^\autom_k) - \Delta(w)
\]
where $\Delta(w)$ is as defined in Algorithm~\ref{algo:optworkload},
with equality achieved for any $w = |\mathcal N|$ by choosing $w$ tasks
with top referral indices.
Thus, choosing an allocation $\mathcal{N}$ that maximizes $\Delta(w)$ is 
equivalent to choosing one that minimizes $J(\mathcal{N}, \{D_k\}_{k \in \mathcal{K}}, p^{\autom}_{1:K})$.
\end{IEEEproof}

%% file: texFiles/simulations.tex
\section{Numerical Simulations}
\label{section: simulations}

In this section, we present a simulation study to compare the performance of the proposed optimal referral policy with two other baseline policies. To run simulation experiments, we need a model for the human operator's decision making. We assume that the human operator has a Gaussian observation model and classifies tasks using the optimal Bayes classification rule. This allows us to derive analytic expressions for the human operator's TPR and FPR functions and use them in the computations. The details of this model are presented in the next section.

\subsection{Human decision model}\label{sec:human-model}

We model the human operator as an optimal Bayesian classifier that chooses between $\mathcal{H}_0$ and $\mathcal{H}_1$ using the Bayes likelihood ratio test~\cite{Vincentpoor:Springer91:detection}. In particular, for tasks $n \in \mathcal{N}$ referred to her, the human operator's decision rule  is
\begin{equation}    \label{eq: decision rule Bayes}
D_{n} = \begin{cases}
\mathcal{H}_0, & \text{ if } p^\human_{1,n}(w) < \rho,  \\
\mathcal{H}_1, & \text{ if } p^\human_{1,n}(w) \geq \rho,
\end{cases}
\end{equation}
where the decision threshold $\rho$ is given by
\[
    \rho = \frac{c_{\tup{fp}}-c_{\tup{tn}}}{c_{\tup{fp}} - c_{\tup{tn}} + c_{\tup{fn}} - c_{\tup{tp}}}.
\]
In our simulations, we will assume that the human operator's observation model is 
a Gaussian observation model as in Example~\ref{ex: Gaussian models general}. 
For such models, the TPR and the FPR functions for the threshold decision rule 
may be written as
\begin{equation*}
P^{\human}_{\tup{fp}}(w) = \Qfun \left(\frac{\tau(w)}{\sigma_{\human}(w)} \right), \\
P^{\human}_{\tup{tp}}(w) = \Qfun \left(\frac{\tau(w)-\mu_{h}(w)}{\sigma_{\human}(w)} \right)
\end{equation*}
where $\Qfun(x):=\frac{1}{\sqrt{2\pi}}\int_x^{\infty} e^{-z^2/2}dz$ denotes
the tail distribution function of the standard normal distribution and
\[
\tau(w) = \frac{\mu_{h}(w)}{2} + \frac{\sigma^2_{\human}(w)}{\mu_{h}(w)} 
\ln \left( \frac{(c_{\tup{fp}}-c_{\tup{tn}})\pi_{0}}{(c_{\tup{fn}}-c_{\tup{tp}})\pi_{1}}  \right).
\]

\subsection{Baseline policies}\label{sec:baseline}

We refer to the policy described in Algorithm~\ref{algo:optworkload} as the 
\textsc{optimal allocation} (OA) policy and compare it with two baseline policies: \textsc{blind allocation} (BA) and \textsc{static allocation} (SA). Both baseline policies serve a precomputed \emph{constant} task load to the human operator, but use different assumptions to compute the task load levels. The motivation for considering such fixed task load policies is that variations in task load across different batches can induce stress and fatigue, degrading human performance~\cite{Hancock:1989:DynamicStress, Hockey:1997:CompensatoryControl, Galy:2012:RelationshipMentalCognitiveWorkload}. So, policies with constant task load 
might have an advantage in practice over OA where task load varies over batches.

\subsubsection{\textsc{Blind allocation} (BA)}
In BA, the automation refers a constant number $w_{\tup{ba}}$ of tasks to the human at each round. The tasks to be referred are chosen uniformly at random without taking the observations $Y^{\autom}_{1:K}$ into account. Let $P^{\autom}_{\tup{tp}}$ and $P^{\autom}_{\tup{fp}}$ denote the TPR and FPR functions of the automation (which do not depend on the workload). The optimal choice of $w_{\tup{ba}}$ is given by
\[
    w_{\tup{ba}} = \arg\min_{w\in\mathcal{W}}\{(K-w) \bar \Gamma^{\autom} + w \bar \Gamma^h(w)\},
\]
where $\bar \Gamma^{\autom}$ and $\bar \Gamma^{\human}(w)$ are the expected cost of referring a randomly selected task respectively to the automation and human (when the human's task load is $w$) and these costs are evaluated as follows:
{\small{
\begin{align*}
    \bar \Gamma^{\autom} = \pi_{1} &\bigl[ P^{\autom}_{\tup{tp}} \, c_{\tup{tp}} 
    + (1-P^{\autom}_{\tup{tp}}) \, c_{\tup{fn}} \bigr] + \pi_{0} \bigl[ P^{\autom}_{\tup{fp}} \, c_{\tup{fp}} + (1-P^{\autom}_{\tup{fp}}) \, c_{\tup{tn}} \bigr],\\
    \bar \Gamma^{\human}(w) &= c_r + \pi_{1} \bigl[ P^{\human}_{\tup{tp}}(w) \, c_{\tup{tp}} + (1-P^{\human}_{\tup{tp}}(w)) \, c_{\tup{fn}} \bigr] \nonumber\\ 
    &\quad + \pi_{0} \bigl[ P^{\human}_{\tup{fp}}(w) \, c_{\tup{fp}} + 
    (1-P^{\human}_{\tup{fp}}(w)) \, c_{\tup{tn}}\bigr].
\end{align*}
}}
\subsubsection{\textsc{Static allocation} (SA)} Similar to BA, in SA the automation refers a constant number $w_{\tup{sa}}$ of tasks to the human in each round. However, the tasks to be referred are chosen after taking the observations $Y^{\autom}_{1:K}$. For a given $Y^{\autom}_{1:K}$ (i.e., a given $p^{\autom}_{1:K}$) let $\bar J(w, p^{\autom}_{1:K})$ denote the minimum value of $J(\mathcal{N}, \{D_k\}_{k \in \mathcal{K}\setminus\mathcal{N}}, p^{\autom}_{1:K})$ for a given task load $w = |\mathcal{N}|$ (obtained by the solution proposed in Theorem~\ref{thm:fixed_w}).  The optimal choice $w_{\tup{sa}}$ is given by
\begin{equation}    \label{eq:w-sa}
    w_{\tup{sa}} = \arg\min_{w \in \mathcal{W}} \mathbb{E}_{p^{\autom}_{1:K}} \bigl[ \bar J(w, p^{\autom}_{1:K}) \bigr].
\end{equation}
Note that the above is a stochastic optimization problem, which can be solved numerically when the distribution of $p^{\autom}_{1:K}$ is known, as is the case in this simulation study. However, if the SA policy is to be used in real-world experiments involving human participants, the above optimization problem would need to be solved using sampling based methods, which require a large dataset of samples from multiple batches and observing the performance across different task loads. 

\subsection{Simulation setup and comparison of policy performance}

We conduct a simulation study in which we measure the performance of OA, BA, and SA using Monte Carlo simulations. To do so, we consider a parameterized problem instance where some parameters are kept constant, while others are sampled independently according to specific distributions. A problem instance is described as follows:
\begin{itemize}[leftmargin=*]
    \item  The hypotheses are chosen with $\pi_0 = 0.8$ (and $\pi_1 = 0.2$). 
    \item The observation model of the human operator and the automation are modeled using the Gaussian observation model of Example~\ref{ex: Gaussian models general} - case 1. The automation's observation model is parameterized by $(d_\circ, \sigma_a)$ and the human's observation model is parameterized by $(\mu_\circ, \sigma_\circ)$. These parameters are chosen according to $d_\circ = \mu_\circ = 3$ and $\sigma_a \sim U(1.5, 2)$ and $\sigma_\circ \sim U(1, 1.5)$. 
    \item  The decision costs are chosen according to $c_{\tup{fp}}, c_{\tup{fn}} \sim U(8,12)$, $c_{\tup{tp}}, c_{\tup{tn}} \sim U(0,2)$, and $c_r \sim U(0, 0.5)$. 
\end{itemize}
The automation refers a subset of tasks to the human according to the OA, BA, or SA policies. For the remaining tasks, it chooses a classification decision according to~\eqref{eqn:auto_Bayes_decision}. We assume that the tasks referred to the human are classified according to the optimal Bayes classifier as described in Sec.~\ref{sec:human-model}.

\begin{figure}[!h]
\begin{tabular}{cc}
    \centering
        \hspace{-0.27cm}\includegraphics[width=0.51\columnwidth]{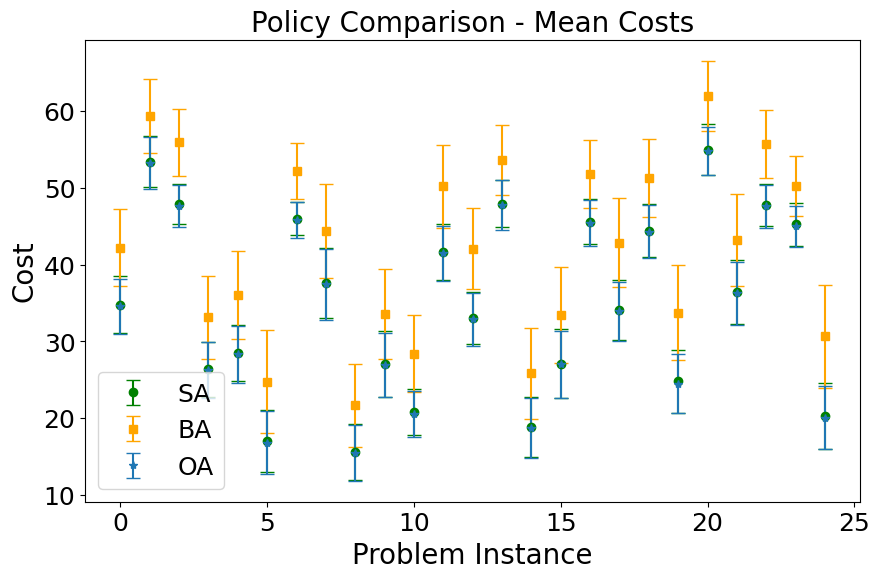}
        & \hspace{-0.4cm}\includegraphics[width=0.5\columnwidth]{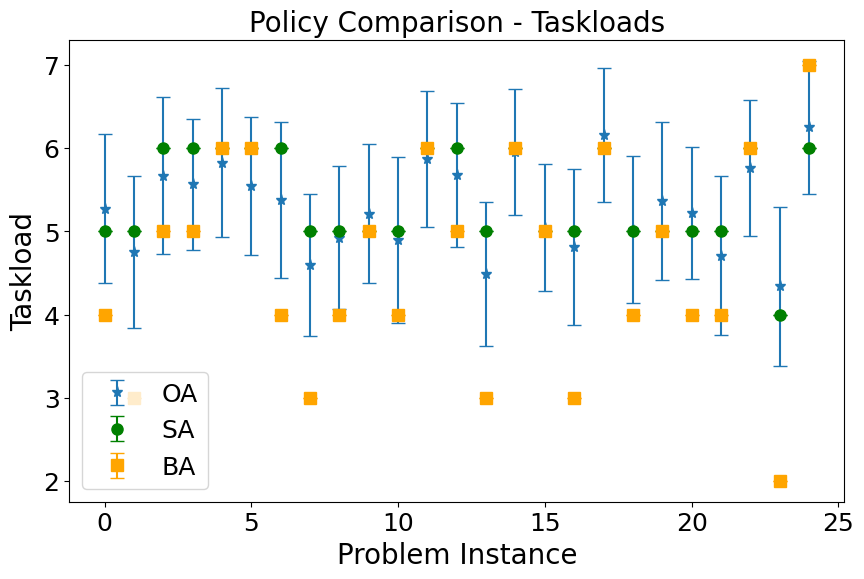}
\end{tabular}
    \caption{Comparison of OA, BA, and SA for $25$ randomly sampled problem instances. For each instance, the figures show summary statistics when the experiment is repeated $2000$ times.}
    \label{fig:Compare_25_PIs}
\end{figure}

We consider a randomized study in which we generate $25$ problem instances where we sample the parameters
$\{\sigma_a, \sigma_h, c_{\tup{fp}},c_{\tup{fn}},c_{\tup{tp}},c_{\tup{tn}},c_{\tup{r}}\}$ as described above.
For each problem instance,  we generate $2000$ 
batches of tasks of size $K=20$. The average performance of the OA, SA and BA policies (over the $2000$ batches) is 
summarized in Fig.~\ref{fig:Compare_25_PIs}. 

Fig.~\ref{fig:Compare_25_PIs}[\emph{left}] shows the mean and standard deviation of the average performance of the three policies for each problem instance. It can be seen from the plots that OA performs much better than BA for every problem instance (in particular, the cost of OA is  at least $15\%$ smaller and its standard deviation at least $3\%$ smaller than those of BA) while the OA and SA have almost the same performance. 

Fig.~\ref{fig:Compare_25_PIs}[\emph{right}] shows the task load of the human operator under the three policies. Recall that for a specific problem instance, BA and SA allocate the same number of tasks to the human operator across different batches.  
We also observe that for this simulation study, OA shows a relatively small variation 
in task load,
which can explain the small performance gap observed here between OA and SA. 
Under conditions where the optimal policy would result in larger task load variations, 
a larger gap in performance could be expected.

%% file: texFiles/experiments_human.tex
\section{Experimental study with human participants}
\label{sec:humans}

In this section we evaluate the practical effectiveness of the proposed referral policy using an experimental study with voluntary human participants. The study comprises two experiments. In the first experiment, we estimate the TPR and FPR of human operators (as defined in~\eqref{eq: S2 decision probabilities}) as functions of task load. As an additional point of interest, we show that these estimates are well approximated by a simple randomized decision model for the human. In the second experiment, we use these estimates to compare the performance of the OA policy (Algorithm~\ref{algo:optworkload}) and  the BA policy defined in Section~\ref{sec:baseline}. 

\subsection{Description of the Experiment}

\begin{figure}[!ht]
\centering
\includegraphics[width=\columnwidth]{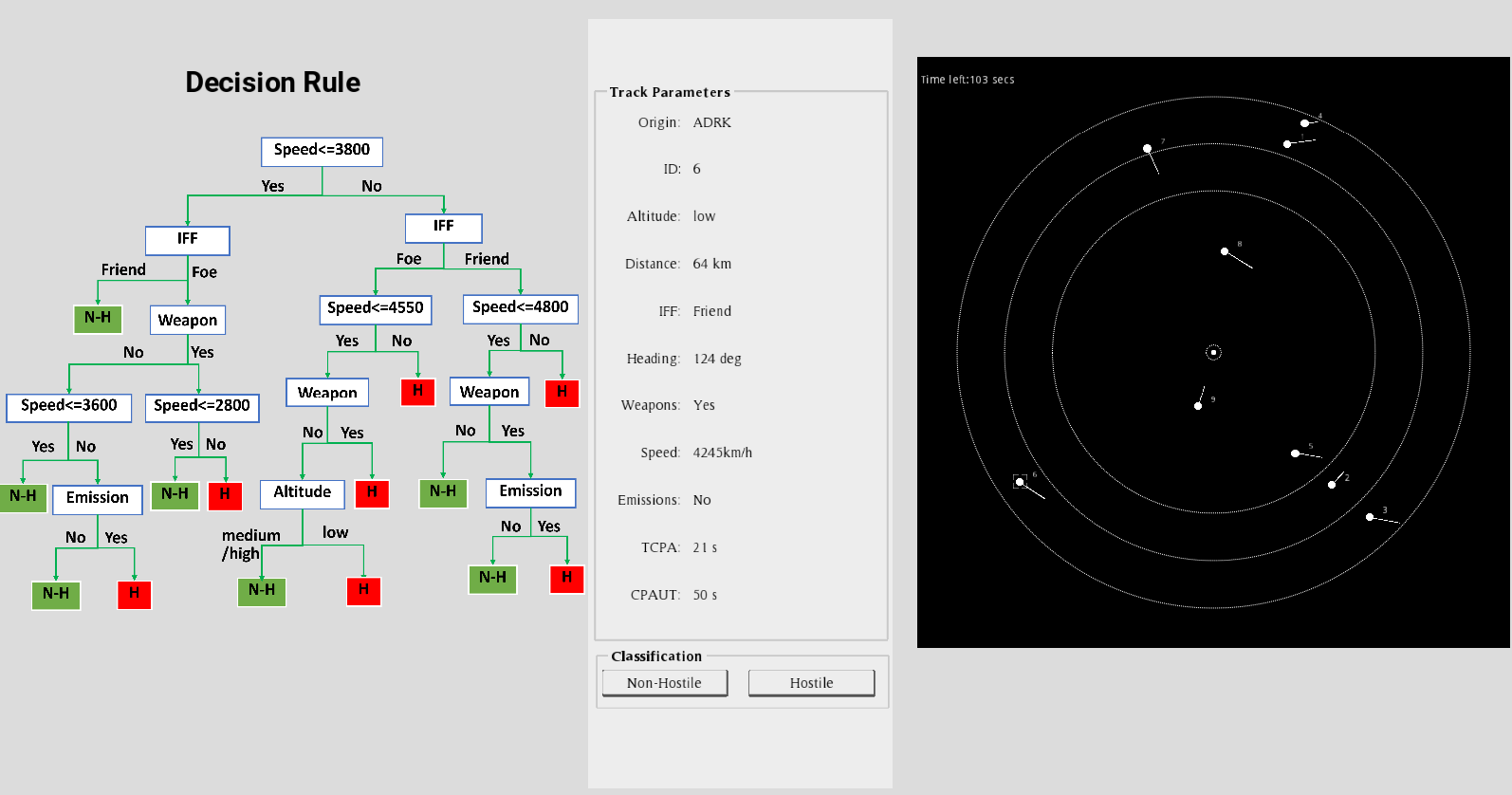}
\caption{Experimental setup: A simulation of a radar screen showing multiple mobile objects. 
Clicking an object displays information such as speed, altitude, etc, on the central pane. 
The operator tries to follow the decision tree to decide the correct label (\emph{hostile} 
or \emph{non-hostile}) and presses one of the two classification buttons.}
\label{fig:sccs}
\end{figure}

Our study uses a simulation environment for threat evaluation closely imitating the S-CCS microworld 
(Simulated Combat Control System) described in \cite{Lafond:HF09:SCCS}.
Participants monitor a simulated radar display, shown on Fig.~\ref{fig:sccs},
with a set of mobile objects represented as dots on the screen, each of which is 
either \emph{hostile} ($\Hyp_1$) or \emph{non-hostile} ($\Hyp_0$). 
Upon clicking an object, information such as speed, altitude, origin, distance, 
direction, weapons, electronic emission signature, identification response, etc., 
is displayed. These parameters are generated with different probability distributions 
for hostile and non-hostile objects. 

The participants are asked to classify all the objects on the screen as hostile or non-hostile 
by following a decision tree (shown in Fig.~\ref{fig:sccs}) provided to  them.  This tree specifies thresholds or intervals  for various parameters to reach a classification decision. 
Correctly following the decision tree yields a TPR of $87\%$ and an FPR of $4.6\%$.
However, at higher task loads the operators 
might make mistakes when following the tree, skip part of it,
or be unable to classify all objects,
thereby reducing the accuracy. 

We perform two experiments (with separate groups of participants) to the compare OA and BA policies. Both policies rely on the human operator's TPR and FPR from \eqref{eq: S2 decision probabilities}, which depend on the task load. In preliminary experiments it was observed that the average time taken to complete a classification task by following the decision tree varies between $9$ and $15$ seconds, which corresponds to $8$ to $13$ tasks in 2 minutes. As a result, we restrict the feasible task loads to a slightly larger range $\mathcal{W} = \{6, 7, \dots, 15\}$.

The purpose of the first experiment is to estimate human operator's TPR and FPR for each workload.
To keep the duration of the experiment session reasonable, we estimate human operator's TPR and FPR for four equally spaced task loads in $\mathcal{W}$, i.e., for $w \in  \mathcal{W}_{\mathrm{est}} = \{6, 9, 12, 15\}$. 
We then use linear interpolation to estimate human operator's TPR and FPR for all task loads in
$\mathcal{W}\setminus \mathcal{W}_{\mathrm{est}}$. 
The purpose of the second experiment is to compare the performance of OA and BA policies designed based on human operator's TPR and FPR estimated in the first experiment.

\subsubsection*{Participant Group}
Adult participants (aged $18$ and above) were recruited through public advertisements for 
the experimental study, and received a compensation of CAD 25 for their participation. 
Informed consent was obtained from each participant prior to the start of the experiment. 
The two experiments were conducted in sequence with separate groups of participants.
The first experiment had $N_s = 28$ participants with $19$ males and $9$ females with a median age 
of $26$ years. The second experiment had $N_s = 15$ participants with $7$ males and $8$ females with a median age of $25$ years. Data from participants who completed less than $55\%$ of the total number of tasks were considered invalid and discarded, giving $N_p=18$ valid samples for the first experiment and $N_p=14$ valid samples for the second experiment.

\subsection{Experiment 1: Estimation of Human TPR and FPR} 
\label{section: model fitting}

The purpose of the first experiment was to estimate human operator's TPR and FPR. 
The experiment started with $3$ practice rounds followed by $24$ rounds of two minutes each. The practice rounds were for the participants to become familiar with the experimental setup.
In the next $24$ rounds, the participants were given a randomly chosen task load from the set $\mathcal{W}_{\mathrm{est}} = \{6, 9, 12, 15\}$, where the randomization was done in such a way that each task load was chosen six times. The tasks in each round were randomly generated to require the human operator to navigate the decision tree (shown in Fig.~\ref{fig:sccs}) to a depth of 4 or 5 levels, ensuring similar task complexity across batches and participants.
After each round, the participants were prompted to rest as needed to prevent fatigue before proceeding to the next round by clicking the ``Next'' button on the screen.
Resting between rounds ensured that fatigue did not influence performance so that the data collected across rounds were independent. At the end of each round, tasks left unclassified by the human were randomly classified by the system as \emph{hostile} or \emph{non-hostile} with approximately 50\% probability for each outcome.\footnote{This choice was made for convenience. There are other ways to assign outcomes to unclassified objects and for a specific application a different method can be chosen with corresponding changes in the computation of TPR and FPR functions.}

\begin{figure}[htbp]
  \begin{tabular}{ll}
    \hspace{-0.3cm}
    \includegraphics[width=0.49\columnwidth]{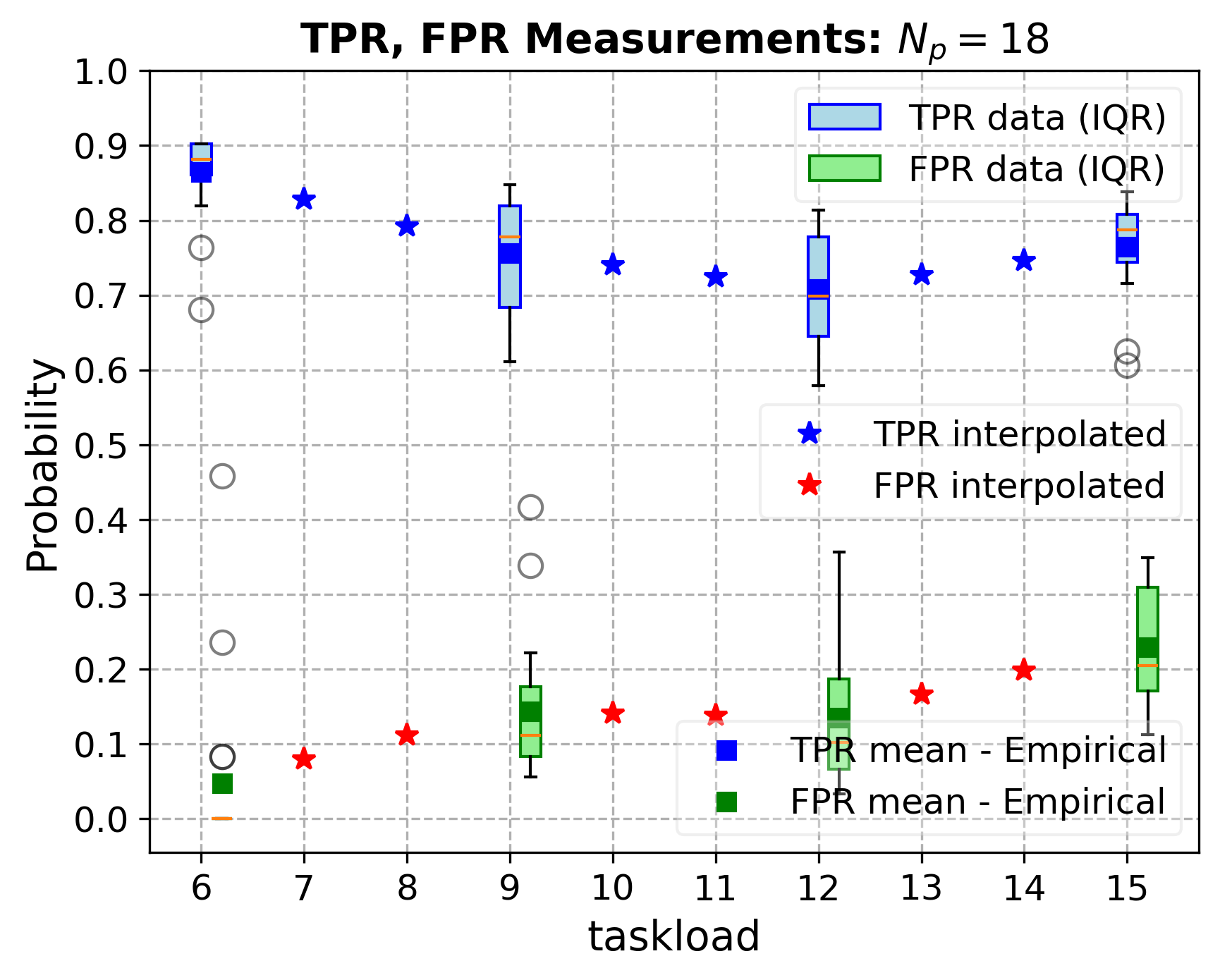}
  &
    \hspace{-0.4cm}
    \includegraphics[width=0.49\columnwidth]{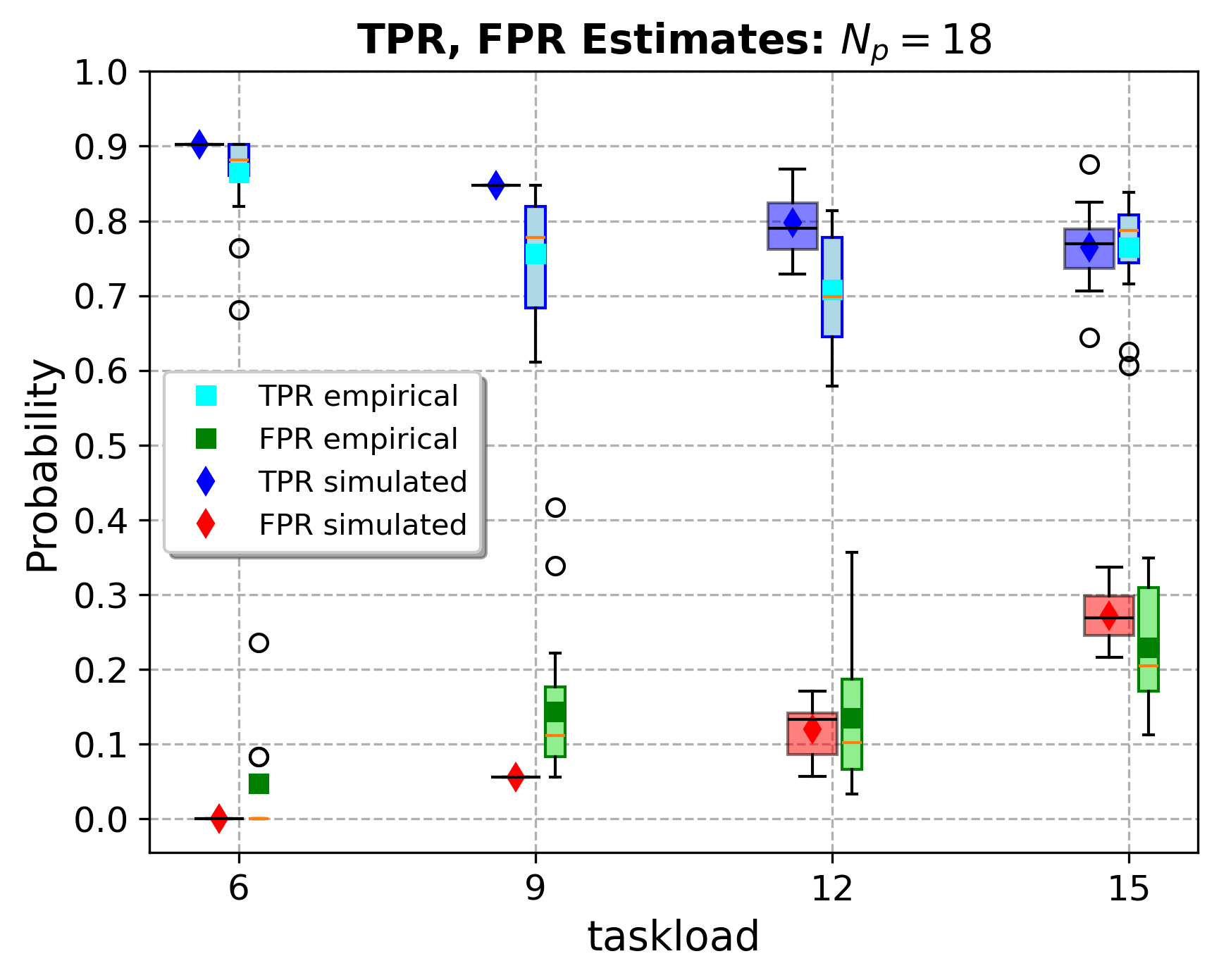}
   
  \end{tabular}
 
  \caption{Experiment 1: [\emph{left}] TPR and FPR measurements for the valid {$18$ participants} 
  (who completed at least $55\%$ of all allocated tasks)
  for four task loads $6, 9, 12, 15$, with inter-quartile (IQR) ranges, along with the outliers represented as circles, and linear interpolation of TPR and FPR for other task loads. [\emph{right}] Comparison of the empirical estimates  
  at the four measured task loads  
  against simulation based estimates  made prior to the experiment.} 
  \label{fig:Experiment1_results}
\end{figure}

At the end of the experiment, the empirical averages of $P^{\human}_{{tp}}(w)$ and $P^{\human}_{{fp}}(w)$ were computed for each task load $w \in \mathcal{W}_{\mathrm{est}}$ using measurements from all $N_p = 18$ samples. Then, estimates for other $w\in\mathcal{W} \setminus \mathcal{W}_{\mathrm{est}}$ were computed using linear interpolation. The results are summarized in Fig.~\ref{fig:Experiment1_results}[\emph{left}], which shows a box and whisker plot of the TPR and FPR measurements, as well as the resulting
interpolation.

Fig.~\ref{fig:Experiment1_results}[\emph{right}] shows that the empirical human operator's 
TPR and FPR values match relatively well the values obtained from the following randomized 
decision model: assume that the human operator takes $12$ seconds\footnote{Recall that in preliminary experiments, we had observed that the time taken to complete a classification task by following the decision tree takes $9$ to $15$ seconds; the chosen value of $12$ second is the average of this range.} to complete each task. For task loads smaller than $10$ (i.e., $120$ seconds divided by $12$ seconds per task), the human operator classifies all tasks using the decision tree. For task loads larger than $10$, the human operator completes $10$ of the tasks (picked randomly withing the assigned tasks) by following the decision tree; the remaining tasks are then randomly classified as \emph{hostile} or \emph{non-hostile} (with approximately $50\%$ probability for each outcome) by the system. The TPR and FPR of this synthetic model is within an error tolerance of $0.1$ from those obtained by the empirical experiment, which is a reasonable error tolerance as suggested by separate numerical simulations performed as in Section~\ref{section: simulations}.

\subsection{Experiment 2: Optimal Allocation versus Blind Allocation}

\begin{figure}[!ht]
    \centering
    \begin{tabular}{c c}
    \hspace{-0.45cm}
    \raisebox{0.45cm}{\includegraphics[width=0.48\columnwidth]{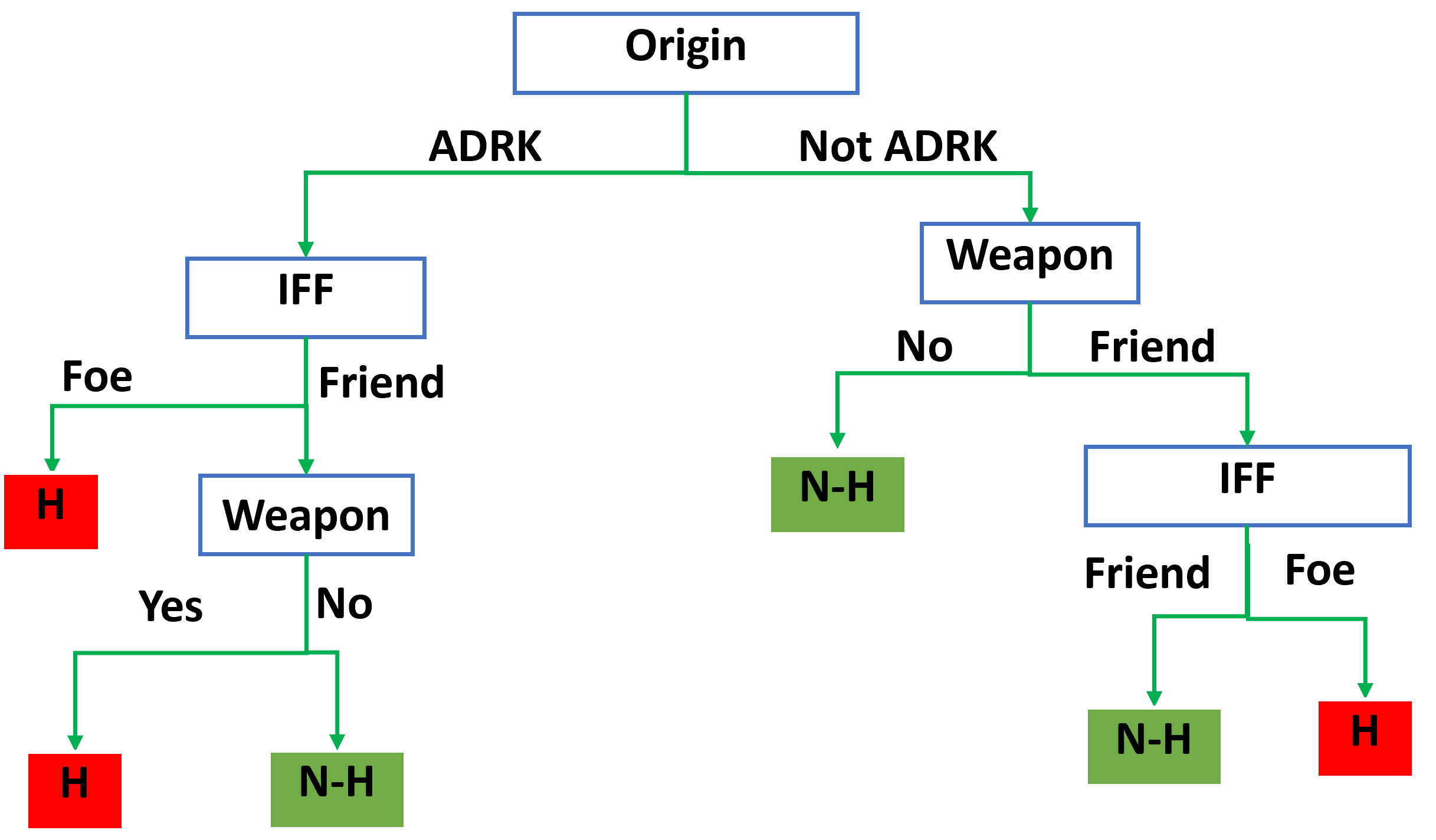}}
    &\hspace{-0.45cm}
    \includegraphics[width=0.52\columnwidth]{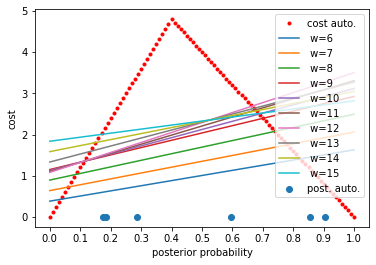}
    \end{tabular}
\caption{[\emph{left}] The decision tree used by the automation to make classification decisions. The referral decisions are made based on posterior probabilities of the classification tasks, computed using the frequency of visiting each of the leaves in the decision tree.
[\emph{right}] Classification decision cost as a function of the posterior probability $p^a_k$ for the experimental scenario, based on the empirically estimated TPR and FPR. The blue dots represent the average posterior probabilities of tasks corresponding to each of the six leaves in the decision tree of the automation. 
}
\label{fig:Decision_tree_auto_and_Cost_vs_posterior_exp}
\end{figure}

The purpose of the second experiment was to compare the performance of OA and BA policies using a ``within-subject'' experiment with a separate group of participants who did not participate in the first experiment. To do so, we generated multiple batches of tasks; for each batch we used both the BA and OA policies to determine a subset of tasks to be referred to the human operator. The tasks not referred to the human operator were classified by the automation by following a different decision tree, shown in Fig.~\ref{fig:Decision_tree_auto_and_Cost_vs_posterior_exp}{[\emph{left}]}. 
To ensure that the accuracy of the human operator at low task load was higher
than the automation's, the latter made its decisions 
by following a decision tree  that was simpler than the decision tree of the human operator 
and combined multiple leaf nodes of the human operator's decision tree into a single node. 
In practice, this could correspond for instance to a situation where the automation 
has access to less information than the human operator.
The automation's decision tree had 
a TPR of $81\%$ and an FPR of $18\%$, to compare to the TPR of 87\% and 
FPR of 4.6\% of the human operator's decision tree.

Experiment~2 followed the same structure as Experiment~1, beginning with three practice rounds followed by 24 two-minute rounds. Before the start of the experiment, $12$ batches of size $30$ tasks were generated, ensuring that $5$ tasks corresponded to each of the $6$ leaves of the automation's decision tree (and corresponded to depth $4$ or $5$ in the human operator's decision tree). The BA and OA policies were computed using decision and referral costs
{{
\begin{equation}\label{eq:exp2-cost}
    c_{\tup{fp}}=8, \quad  c_{\tup{fn}}=12, \quad c_{\tup{tp}}=0, \quad c_{\tup{tn}}=0, \quad c_r=0,
\end{equation} }}
as well as the TPR and FPR functions estimated in Experiment~1 (shown in Fig.~\ref{fig:Experiment1_results}).
The corresponding classification costs are shown in Fig.~\ref{fig:Decision_tree_auto_and_Cost_vs_posterior_exp}{[\emph{right}]}, following
the same principles as for Fig.~\ref{fig:Ex1_C1_EC2_w}.

The BA and OA policies were then applied separately to each batch to determine which tasks were referred to the human operator, generating the set of tasks for the human operator for the $24$ rounds, which were presented in a random order. 
The BA policy resulted in a task load of $w^*_{\mathrm{ba}} = 7$, resulting in $84$ tasks being referred to the human across $12$ BA rounds. Meanwhile, the OA policy's task load rarely exceeded $w = 9$, resulting in $70$ to $100$ tasks being referred to the human across $12$ OA rounds. 

Participants were instructed to complete all assigned tasks using the provided decision tree,  but were not informed of the decision and referral costs. 
The TPR and FPR functions in Experiment~2 were computed and found to be comparable to those of Experiment~1, indicating that the participants in both experiments had similar distribution.

We define the performance of BA and OA policies as the cumulative cost $\sum_{k \in \mathcal{K}} C(D_k, H_k)$ computed using the decision costs in~\eqref{eq:exp2-cost}. We denote these performances by $\mu_{\tup{BA}}$ and $\mu_{\tup{OA}}$, respectively.
For each subject, we have $12$ measurements of $\mu_{\tup{BA}}$ and $\mu_{\tup{OA}}$. Let $\bar \mu_{\tup{BA}}$ and $\sigma_{\tup{BA}}$ denote the mean and standard deviation of the BA policy. Similar notation is used for the OA policy.

\begin{figure}[!h]
    \centering
    \includegraphics[width=0.58\columnwidth]{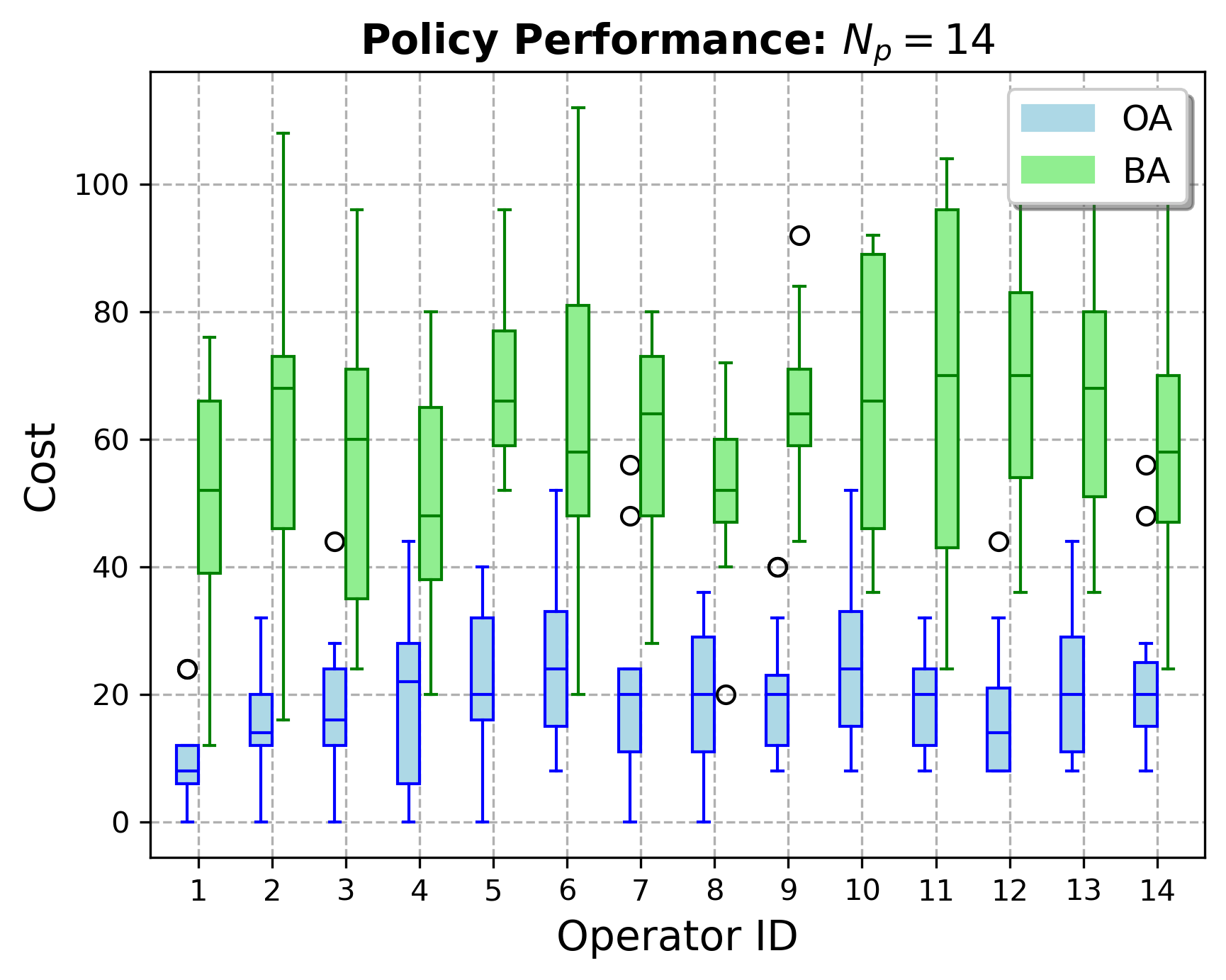}
    \caption{Experiment 2 - Performance measurements (inter-quartile ranges) of 
    BA and OA
    policies for all participants.}
    \label{fig:Exp2_costs_blind_optimal}
\end{figure}

The average and summary statistics for each subject are shown in Fig.~\ref{fig:Exp2_costs_blind_optimal}. 
To determine whether there is a statistically significant performance difference between the two policies, we conduct a paired $t$-test. This test assumes that 
the subjects are independent and that the difference in the performance of both policies is normally distributed~\cite{Montgomery:book2017:DAE}. 
The paired $t$-test is used to evaluate the following hypothesis:
\begin{align*}
    \mbox{Null Hypothesis: } \bar{\mu}_{\tup{BA}} -\bar{\mu}_{\tup{OA}}=0,\\
    \mbox{Alternate Hypothesis: } \bar{\mu}_{\tup{BA}}-\bar{\mu}_{\tup{OA}}> 0.
\end{align*}
The number of degrees of freedom $df$ for the $t$-test is one less than the total number of measurement pairs. Thus, in our case   $\tup{df} = N_{p}-1=13$.

We compare both the average-case and the worst-case performance. The average-case performance compares $\bar \mu_{\tup{BA}}$ and $\bar \mu_{\tup{OA}}$, while the worst-case performance compares $\bar \mu_{\tup{BA}} + \sigma_{\tup{BA}}$ (upper confidence bound on the average performance) with $\bar \mu_{\tup{OA}} - \sigma_{\tup{OA}}$ (lower confidence bound on the average performance). The results are as follows:
\begin{enumerate}[leftmargin=*]
    \item \emph{Average-case analysis:} A paired $t$-test for the average-case performance gave a $t$-statistic of $25.61$ with a $p$-value of $1.6 \times 10^{-12}$.
    \item \emph{Worst-case analysis:} A paired $t$-test for the worst-case performance gave a $t$-statistic of $4.99$ with a $p$-value of $2 \times 10^{-4}$.
\end{enumerate}
In both cases, a large $t$-statistic suggests that there is a substantial difference between the performance of OA compared to the performance of BA. The small $p$-values indicate that the likelihood of observing this result under the null hypothesis is practically zero.
Hence, we reject the null hypothesis that both policies have the same performance 
and accept the alternate hypothesis that the OA policy on average achieves a better performance than the BA policy.

